\documentclass[twocolumn,showpacs,showkeys,preprintnumbers,superscriptaddress,amsmath,floatfix,amssymb,secnumarabic,nofootinbib]{revtex4}
\usepackage[colorlinks=true]{hyperref}
\usepackage{graphicx}
\usepackage{braket}
\usepackage{amsmath}
\usepackage{epsfig}
\usepackage{float}
\usepackage{caption}
\usepackage{subcaption}

\def\({\left(}
\def\){\right)}
\def\[{\left[}
\def\]{\right]}

\newcommand{\Tr}{ {\rm Tr} \, }

\newcommand{\beq} {\begin{eqnarray}}
\newcommand{\eeq} {\end{eqnarray}}

\newcommand{\comment}[1]{}

\begin{document}
\sloppy

\title{Many-body effects in graphene beyond the Dirac model with Coulomb interaction}

\author{N.~Yu.~Astrakhantsev}
\email{nikita.astrakhantsev@itep.ru}
\affiliation{Institute for Theoretical and Experimental Physics, Moscow, 117218 Russia}
\affiliation{Moscow Institute of Physics and Technology, Dolgoprudny, 141700 Russia}

\author{V.~V.~Braguta}
\affiliation{Institute for High Energy Physics NRC "Kurchatov Institute", Protvino, 142281 Russian Federation}
\affiliation{Far Eastern Federal University,  School of Biomedicine, 690950 Vladivostok, Russia}

\author{M.~I.~Katsnelson}
\affiliation{Radboud University, Institute for Molecules and Materials,
Heyendaalseweg 135, NL-6525AJ Nijmegen, The Netherlands}
\affiliation{Ural Federal University, Theoretical Physics and Applied Mathematics Department, Mira Str. 19,  620002 Ekaterinburg, Russia}

\begin{abstract}

This paper is devoted to development of perturbation theory for studying the properties of graphene
sheet of finite size, at nonzero temperature and chemical potential. The perturbation theory is based on the tight-binding
Hamiltonian and arbitrary interaction potential between electrons, which is considered as a perturbation.
One-loop corrections to the electron propagator and to the interaction potential at nonzero temperature
and chemical potential are calculated.
One-loop formulas for the energy spectrum of electrons in graphene, for the renormalized Fermi velocity and
also for the dielectric permittivity are derived.

\end{abstract}

\maketitle

\section{Introduction}
\label{IntroductionSec}

Graphene is a two dimensional crystal composed of carbon atoms which are packed in a honeycomb (hexagonal) lattice \cite{Novoselov:04:1, Geim:07:1}. It attracts considerable interest because of its unique electronic properties; most of them are related to existence of two conical points in the electron energy spectrum (Fermi points) and ``massless Dirac fermion'' character of charge carriers with energy and momentum close to the Fermi points \cite{wallace,mcclure,semenoff,Novoselov:05:1,zhang}. It results in numerous quantum relativistic phenomena such as Klein tunneling, minimal conductivity through evanescent waves, relativistic collapse at a supercritical charge, etc., establishing an interesting and fruitful relation between fundamental physics and materials science \cite{Novoselov:07:1,Beenakker,Novoselov:09:1,Guinea,Kotov:2,Katsnelson}. The effective ``velocity of light'' (Fermi velocity) for the Dirac fermions in graphene is relatively small, $v_F \sim c/300$, and the interaction between the quasiparticles in graphene can be approximated by the instantaneous Coulomb potential with the effective coupling constant\footnote{In this paper we work in units $\hbar = 1$} $\alpha_{eff}\sim \alpha (c /v_F) \sim 300/137 \sim 2$. This interaction is therefore quite strong, which results in a rich variety of phenomena \cite{Kotov:2}. Within the Dirac model, before the experimental discovery of graphene, it was shown  that the long-range character of Coulomb interaction results in a renormalization of the Fermi velocity which is divergent at zero temperature and zero doping leading to a non-Fermi-liquid behaviour \cite{Gonzalez:1993uz}; this prediction has been recently confirmed experimentally \cite{Elias,PNAS}.

At the same time, Dirac model gives us many-body renormalization of electronic properties only for small coupling constant and only with the logarithmic accuracy. The higher-order terms were considered in Refs. \cite{fogler,dassarma} but still within the Dirac model. To calculate quantitatively correctly these properties one needs to work with the lattice model and with a realistic potential of electron-electron interaction $V_{xy}$ taking into account its screening by the $\sigma$-electrons.  The corresponding first-principle results \cite{Wehling} can be parametrized by the phenomenological potential. This modification of the interaction potential at small scales in comparison with the bare Coulomb potential was proved to affect some graphene properties significantly (for instance, the phase diagram of graphene \cite{Ulybyshev:2013swa}).

The authors of papers \cite{Ulybyshev:2013swa, Buividovich:2012nx, Smith:2014tha} carried out the Monte-Carlo study of graphene properties based on the tight-binding Hamiltonian without the expansion near the Fermi points. Within this approach one can introduce an arbitrary phenomenological potential $V_{xy}$. Using the tight-binding Hamiltonian
on the hexagonal lattice, we are going to build the perturbation theory in $V_{xy}$. We believe that the theory built in this way is
important and interesting for the following reasons.

The theory based on the tight-binding Hamiltonian has more common features with real graphene physics than the effective theory based on
the expansion in the vicinity of Dirac points. For instance, the tight-binding Hamiltonian
``remembers'' graphene properties such as geometry of hexagonal lattice or the
natural energy scale such as $\pi$-bandwidth, which are absent in the effective Dirac theory. Moreover, one can include a phenomenological potential in this theory, which is closer to the real graphene physics than the bare Coulomb potential.

In addition, the theory based on the tight-binding Hamiltonian and the phenomenological interaction $V_{xy}$ can be easily improved.
For instance, one can study the effects appearing due to the inclusion of the next-to-the-nearest-neighbour hopping or nonzero chemical potential.
Note that such study cannot be carried out within the Monte-Carlo simulation because of the well known sign problem.
Finally, if the electron properties of other nanomaterials are formulated in terms of the tight-binding Hamiltonian
and the phenomenological potential, it makes no difficulties to apply the results of this paper to study these materials.

Lattice simulation of graphene were proved to be a very efficient and quickly developing approach for studying the properties of graphene
\cite{Ulybyshev:2013swa, Buividovich:2012nx, Smith:2014tha, Lahde:09, Hands:10:1, Buividovich:2012uk}. An important feature of all these simulations is that they are conducted at the finite lattice, at the finite temperature and with the
finite discretization errors. In order to check the lattice results and estimate the discretization uncertainty in the weak coupling region it is very useful to develop the perturbation theory which accounts all these effects.

Strictly speaking, the theory with the arbitrary phenomenological potential is not renormalizable since it contains four-fermion terms.
However, lattice formulation provides the ultraviolet (spacing between carbon atoms) and the infrared
(the finite size of the lattice) regulators. For this reason the theory
on the hexagonal lattice is well defined.

Finally, one should mention that the interaction in graphene is strong, so the application of the perturbation theory is questionable.
However, we believe that even at the one-loop level one can study some important physical effects.
One can also expect that the perturbation theory built in this paper
can well describe graphene many-body effects similarly to the RPA based on effective theory of graphene\cite{dassarma}. 

This paper is organized as follows. In the next section we built the perturbation theory which is based on the tight binding
Hamiltonian and arbitrary interaction potential between electrons. In the section 3 we calculate one-loop corrections to the electron propagator. The section 4 is devoted to the calculation of one-loop corrections
to the interaction potential. Finally, in the last section we discuss and summarize the results of this paper.
\section{ Partition function and Feynman rules }

\subsection{Geometry}
\label{subsec:basic_defs}

\begin{figure}[t!]
        \centering
                \includegraphics[width=0.4\textwidth]{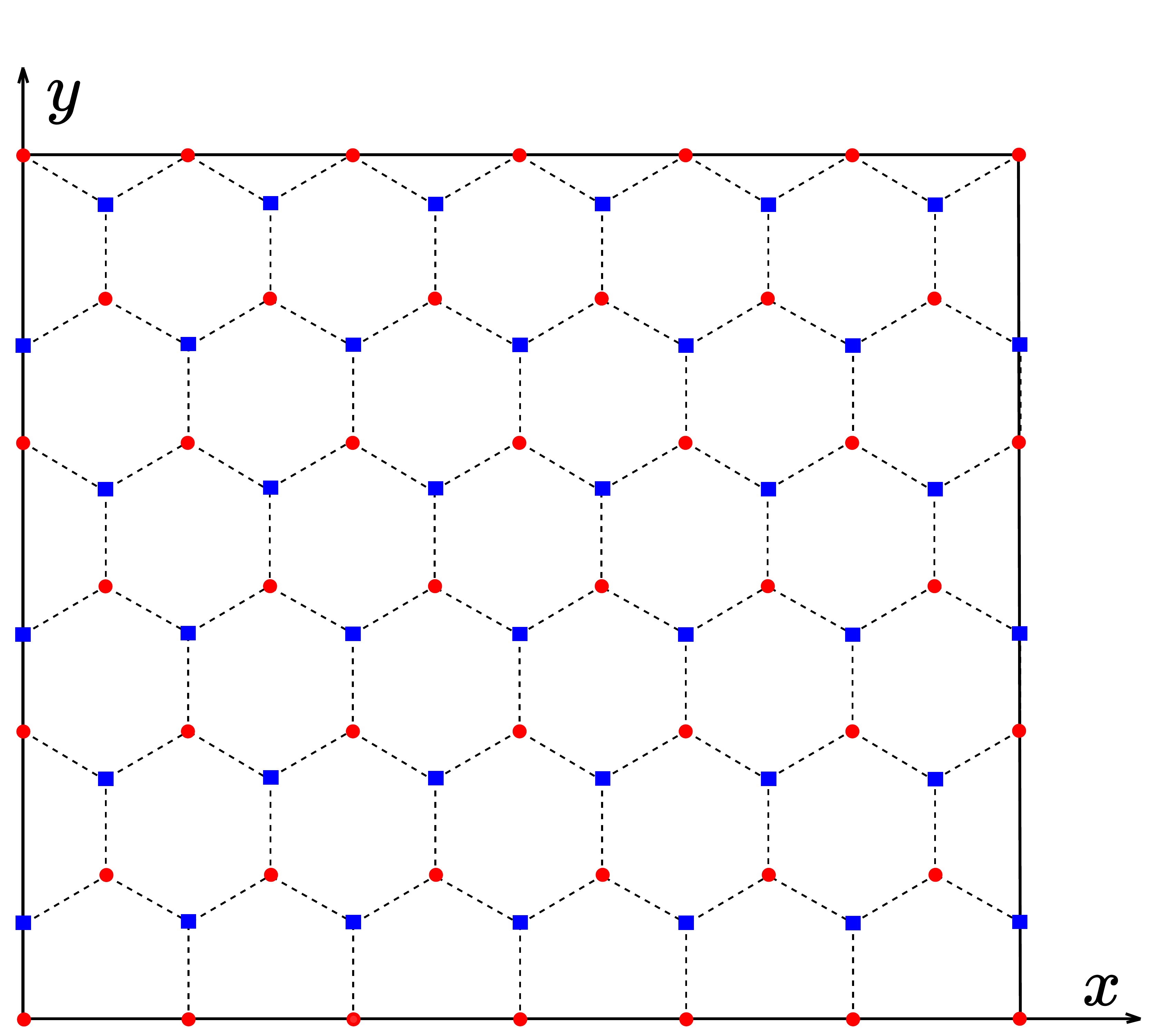}
        \caption{$L_x \times L_y = 6 \times 6$ graphene sheet. The sites belonging to
the sublattices $A$ and $B$ are shown as rectangles and circles respectively. }
        \label{fig:sheet}
\end{figure}

We consider a hexagonal lattice with the torus topology.
The example of such lattice, which consists of $L_x \times L_y = 6 \times 6$ hexagons, is shown in Fig.\ref{fig:sheet}.

The hexagonal lattice is the composition of two triangular sublattices $A$ and $B$. The sites belonging to
the sublattices $A$ and $B$ are shown as rectangles and circles respectively. The Cartesian coordinates $(x, y)$ of any lattice site can be parametrised by three numbers $(s, \xi_1, \xi_2)$, where $s = A / B$ is the sublattice index and $\xi_1 = 0 \ldots L_x-1$, $\xi_2 = 0 \ldots L_y-1$, so
\begin{gather}
x = \sqrt{3} a \xi_1 + \frac{\sqrt{3}}{2} a \xi_2 + \frac{\sqrt{3}}{2} a \delta_{s, B}, \\
y = \frac{3}{2} a \xi_2 - \frac{1}{2} a \delta_{s, B}.
\end{gather}
The torus topology implies the following identification of $\vec{\xi}$ coordinates:
\begin{gather}
(\xi_1 + L_x, \xi_2) \to (\xi_1, \xi_2),\;\; (\xi_1, \xi_2 + L_y) \to (\xi_1 + L_y / 2, \xi_2).
\end{gather}

Every site of the $A$ sublattice is connected to three sites of the sublattice $B$. The vectors
$\vec{\rho}_b$ in $\xi$-coordinates, connect $(\alpha, \vec{\xi})$ to its neighbours $(\beta, \vec{\xi} + \vec{\rho}_b)$:
\begin{gather}
\vec{\rho}_1 = (0, 0), \; \vec{\rho}_2 = (-1, 1), \; \vec{\rho}_3 = (-1, 0).
\end{gather}
In $x$-coordinates these $\vec{\rho}_b$ vectors read:
\begin{gather}
\vec{\rho}_1 = (0, 0), \; \vec{\rho}_2 = \left(-\frac{\sqrt{3}}{2} a, \frac{3}{2} a \right),\; \vec{\rho}_3 = \left(-\sqrt{3} a, 0 \right).
\end{gather}

\subsection{Tight-binding Hamiltonian with interactions}
The electronic properties of graphene can be described by the tight-binding Hamiltonian:
\begin{gather}
\hat{H}_{\mbox{tb}} = - t \sum_{\sigma = \uparrow,\downarrow} \sum_{\Braket{xy}}\left( \hat{c}^{\dagger}_{\sigma, x} \hat{c}_{\sigma, y} + \hat{c}^{\dagger}_{\sigma, y} \hat{c}_{\sigma, x} \right),
\label{hamiltonian}
\end{gather}
where the summation is kept over the neighbouring graphene lattice sites $x$ and $y$, $t \approx 2.7\;\mbox{eV}$ is the hopping parameter. Operators $\hat{c}^{\dagger}_{\sigma, x}$, $\hat{c}_{\sigma, x}$ create and annihilate electron with spin $\sigma=\uparrow,\downarrow$ at the lattice site $x$.
Note that one can also include the next-to-nearest neighbours hopping
to the Hamiltonian (\ref{hamiltonian}), but for simplicity we restrict our consideration to the nearest neighbours.

We choose the vacuum state that satisfies the following conditions:
\begin{gather}
\hat{c}_{\uparrow, x} \Ket{0} = 0,\; \hat{c}^{\dagger}_{\downarrow, x} \Ket{0} = 0,
\end{gather}
so there is an electron with spin $\sigma = \downarrow$ at every lattice site and no electrons with spin $\sigma=\uparrow$. As our main goal is to calculate the partition function $\mathcal{Z}$, which contains summation over all states, the specific choice of $\Ket{0}$ will not affect any physical results.

It is convenient to rewrite the Hamiltonian in terms of creation and annihilation operators for ''particles'' and ''holes'', which read:
\begin{gather}
\hat{a}_{x} = \hat{c}_{+, x},\; \hat{b}_{x} = \pm \hat{c}^{\dagger}_{-, x}.
\end{gather}
The plus sign is taken for $x \in A$ and the minus sign for $x \in B$, where $A$ and $B$ are the triangular sublattices labels.
After operators redefinition the ground state satisfies $\hat{a}_{x} \Ket{0} = \hat{b}_{x} \Ket{0} = 0$. Thereby, we interpret the absence of a valence electron as a positively charged ''hole'' and an additional electron as a negatively charged ''particle''. In terms of these operators the tight-binding Hamiltonian is:

\begin{gather}
\hat{H} = - t \sum\limits_{<xy>} \left(\hat{a}^{\dagger}_{x} \hat{a}_{y} +
\hat{b}^{\dagger}_{y} \hat{b}_{x} + H.c. \right).
\end{gather}

The charge operator now reads:
\begin{gather}
\hat{q}_x = \hat{a}^{\dagger}_{x} \hat{a}_{x} - \hat{b}^{\dagger}_{x} \hat{b}_{x}.
\end{gather}
It easy to check that $\hat{q}_x \Ket{0}  = 0$, which means that this vacuum is electrically neutral.

In some applications \cite{Ulybyshev:2013swa, Buividovich:2012nx} the mass term is introduced:
\begin{gather}
\hat{H}_{m} = \sum\limits_x (\pm m c^2) \hat{a}^{\dagger}_{x} \hat{a}_{x} + \sum\limits_x (\pm m c^2) \hat{b}^{\dagger}_{x} \hat{b}_{x},
\end{gather}
here the plus sign is taken for sublattice A and the minus sign --- for sablattice B . This term explicitly breaks the symmetry between two sublattices.

In order to study the action of the chemical potential on graphene properties we introduce the term
\begin{gather}
\hat{H}_{\mu} =  \mu  \sum\limits_x \bigl (  \hat{a}^{\dagger}_{x} \hat{a}_{x} -    \hat{b}^{\dagger}_{x} \hat{b}_{x} \bigr ).
\end{gather}

It is known that the interaction between electrons plays an important role. This part of the Hamiltonian has the form:
\begin{gather}
\hat{H}_{\mbox{int}} = \frac{1}{2}\sum\limits_{x, y} V_{xy} \hat{q}_x \hat{q}_y.
\end{gather}
The Coulomb potential $V$ is often used to describe the interaction. However, it was shown in \cite{Wehling}, that the real
potential $V_{xy}$ dramatically deviates from the Coulomb law at small distances, which affects physical properties \cite{Ulybyshev:2013swa}.

The aim of this paper is to study the properties of the electronic system, described by the Hamiltonian:
\begin{gather}
\hat{H} = - t \sum\limits_{<xy>} \left(\hat{a}^{\dagger}_{x} \hat{a}_{y} +
\hat{b}^{\dagger}_{y} \hat{b}_{x} + H.c. \right) +  \sum\limits_x (\pm m c^2) \hat{a}^{\dagger}_{x} \hat{a}_{x} + \nonumber\\ +
\sum\limits_x (\pm m c^2) \hat{b}^{\dagger}_{x} \hat{b}_{x} + \mu  \sum\limits_x \bigl (  \hat{a}^{\dagger}_{x} \hat{a}_{x} -    \hat{b}^{\dagger}_{x} \hat{b}_{x} \bigr )
+ \frac{1}{2} \sum\limits_{x, y} V_{xy} \hat{q}_x \hat{q}_y,
\label{eff1}
\end{gather}
where $V_{xy}$ is an arbitrary phenomenological interaction, treated as the perturbation.

\subsection{Electronic spectrum of graphene without interaction}
If the interaction is neglected\footnote{In this subsection we consider graphene with zero chemical potential}, one can easily find Hamiltonian spectrum and its eigenfunctions, which can be written as:
\begin{gather}
\psi_{\vec{k}}^{\zeta}(\vec{x}) = \begin{pmatrix} c^{\zeta}_A (\vec{k}) \\ c^{\zeta}_B(\vec{k}) \end{pmatrix} e^{i \vec{k} \vec{x}},
\end{gather}
where $\zeta = \pm 1$ is an additional label representing particles ($\zeta = + 1$) and antiparticles ($\zeta = -1$). The first vector component corresponds to the $A$ sublattice, similarly the second one corresponds to the $B$ sublattice. The Brillouin zone momentums $\vec{k}$ are:
\begin{gather}
k_x = \frac{2 \pi m_x}{\sqrt{3} a L_x},\; k_y = \frac{2 \pi m_y}{3 a (L_y / 2)}.
\end{gather}
Indices $m_x = 0,\ldots, L_x - 1$, $m_y = 0,\ldots, L_y - 1$ give the full set of the eigenfunctions for the
torus topology of the graphene sheet.

The  energy spectrum of the Hamiltonian is then:
\begin{gather}
\label{phi}
E_{\zeta} (\vec{k}) = \zeta \sqrt{m^2 c^4 +  |\varphi(\vec{k})|^2},\;\;\mbox{where} \;\; \varphi(\vec{k}) = t \sum\limits_{b = 1}^3 e^{i \vec{k} \vec{\rho}_b}.
\end{gather}

The wave function components are
\begin{gather}
c^{\zeta}_A(\vec{k}) = \sqrt{\frac{E(\vec{k}) + \zeta m c^2}{2 E(\vec{k}) L_x L_y}}, \nonumber\\ c^{\zeta}_B(\vec{k}) = -\zeta e^{-i \arg \varphi(\vec{k})} \sqrt{\frac{E(\vec{k}) - \zeta m c^2}{2 E(\vec{k}) L_x L_y}}.
\end{gather}

There exist two special momenta $\vec{k}_F$, such that $\varphi(\vec{k}_F) = 0$ and $E(\vec{k}_F) = \pm m c^2$, called the Dirac points.
There are two different Dirac positioned at: $(m_x, m_y) = (2 L_x / 3, 0)$ and $(L_x / 3, L_y / 2)$.

For the  massless fermions ($m = 0$), in the vicinity of the Dirac points the energy spectrum of the fermion quasiparticles is:
\begin{gather}
E(\vec{k}) = v_F |\vec{k} - \vec{k}_F|, \nonumber \\
v_F^0 = \frac{3}{2} t a \approx \frac c {343} = 0.87 \cdot 10^6~ \frac {m} {s}.
\end{gather}
The linear spectrum of fermion excitations plays the central role in the effective theory of graphene.

\subsection{Partition function}
Let us consider a graphene sheet $L_x \times L_y$ at the temperature $T$. To write down the partition function for this graphene sample, one should handle the 4-fermion interaction operator which is contained in the Hamiltonian (\ref{eff1}). So, before going to the partition function the interaction term of the Hamiltonian should be decomposed using the Hubbard-Stratonovich transformation:
\begin{gather}
\exp \left(-\frac{1}{2} \sum\limits_{x, y} \hat{q}_x V_{xy} \hat{q}_y \right) = \nonumber\\ = \int \mathcal{D} \varphi \exp \left(-\frac{1}{2} \sum\limits_{x, y} \varphi_x V^{-1}_{xy} \varphi_y - i  \sum\limits_x \varphi_x \hat{q}_x \right).
\end{gather}

\begin{figure}[h!]
        \centering
			\includegraphics[width=0.25\textwidth]{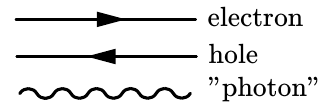}
        		\caption{Graphical representation of the propagators of different particles.}
			\label{propagators}
\end{figure}
In the last expression we introduced the Hubbard field $\varphi$, which carries the interaction. Note also that we
have omitted the determinant of the $V$ which is not important for our calculations.

Now it causes no difficulties to write down the partition function in terms of the path integral.
To this end, we divide the Euclidean time $\tau_E \in [0,\beta],~~\beta=1/T$ into $N_{\tau}$ parts with the size $\delta \tau=\beta/N_{\tau}$.
We introduce the electron fields $(\bar \eta,\eta)$, the hole fields $(\bar \psi,\psi)$ and the
Hubbard field $\varphi$ (which is called ``photon'' below) at each site of the lattice $L_x \times L_y \times \beta$. The boundary conditions
in Euclidean time direction are periodic for ``photon'' $\varphi$  and
antiperiodic for the fermion fields $(\bar \eta,\eta)$, $(\bar \psi,\psi)$.
The boundary conditions for the spatial directions are periodic for all the fields.

Finally, the partition function for the model can be written as \cite{Ulybyshev:2013swa, Buividovich:2012nx}:
\begin{widetext}
\begin{gather}
\label{part_function}
\mathcal{Z} = \int \mathcal{D} \varphi \, \mathcal{D} \bar{\eta} \mathcal{D} \eta \, \mathcal{D} \bar{\psi} \mathcal{D} \psi e^{-\mathcal{S}_{em}(\varphi) - \sum\limits_{\sigma, x, y} \bar{\eta}(x) \mathcal{M}_{x, y} (\varphi) \eta(y) - \sum\limits_{\sigma, x, y} \bar{\psi}(x) \mathcal{\bar{M}}_{x, y} (\varphi) \psi(y)},
\end{gather}
where the matrix $\mathcal{M}_{l_1, l_2, \sigma}(x, y)$ reads (here $l_1 = A, B$ and $l_2 = A, B$ are sublattice indices):
\begin{gather}
\mathcal{M}_{l_1, l_2} (x, y) = \left(
  \delta_{\vec{x},\vec{y}} \delta_{l_1, l_2}  \delta_{x^0, y^0}
- e^{i \delta \tau \varphi_{l_1} (x)} e^{\delta \tau \mu}  \delta_{\vec{x},\vec{y}} \delta_{l_1, l_2} \delta_{x^0 + \delta \tau, y^0}  + t \delta \tau ~ e^{\delta \tau \mu} \left[ e^{i \delta \tau \varphi_{A} (x)} \delta_{l_1, A}, \delta_{l_2, B} \sum\limits_{i = 1}^3 \delta_{x^0 + \delta \tau, y^0} \delta_{\vec{x} + \vec{\rho}_i, \vec{y}} + \right. \right. \nonumber\\  \left. \left. + e^{i \delta \tau \varphi_{B} (x)} \delta_{l_1, B} \delta_{l_2, A} \sum\limits_{i = 1}^3 \delta_{x^0 + \delta \tau, y^0} \delta_{\vec{x} - \vec{\rho}_i, \vec{y}} \right] + m c^2 \delta \tau~ e^{\delta \tau \mu} \left[ e^{i \delta \tau \varphi_{A} (x)} \delta_{l_1, A} \delta_{l_2, A} \delta_{\vec{x}, \vec{y}} \delta_{x^0 + \delta \tau, y^0} - e^{i \delta \tau \varphi_{B} (x)} \delta_{l_1, B} \delta_{l_2, B} \delta_{\vec{x}, \vec{y}} \delta_{x^0 + \delta \tau, y^0}\right] \right), \\
\mathcal{S}_{em}(\varphi) = \frac{\delta \tau}{2} \sum\limits_{x, y} \varphi(x) V^{-1}_{xy} \varphi(y).
\label{Sem}
\end{gather}
\end{widetext}

To get the expression for the fermion operator $\bar M$ one should take the formula for the $M$ and
carry out the substitution $\varphi \to -\varphi$, $\mu \to -\mu$.
Note that in the formula (\ref{Sem}) the summation is taken over all the coordinates $(\xi_1,\xi_2)$ and over the sublattice indices, so the interaction potential $V_{xy}$ is assumed to be a matrix
\begin{gather}
V_{xy} = \begin{pmatrix} V^{AA}(x, y) & V^{AB}(x, y) \\ V^{BA}(x, y) & V^{BB}(x, y) \end{pmatrix}.
\end{gather}
In the limit of $\delta  \tau \to 0$ expression (\ref{part_function}) corresponds to the partition function of graphene sheet $L_x \times L_y$
at the temperature $T$. In some applications, for instance in the Monte-Carlo simulation of graphene \cite{Ulybyshev:2013swa, Buividovich:2012nx},
one uses small but finite step $\delta \tau$. Some of the formulas below are written for the finite $\delta \tau$ to have the possibility to estimate the discretization uncertainty.

\subsection{Propagators}
Using the expression for the partition function (\ref{part_function}), one can write down free propagators for the corresponding fields
(see Fig. \ref{propagators}).

\begin{figure}[t!]
        \centering
                \includegraphics[width=0.25\textwidth]{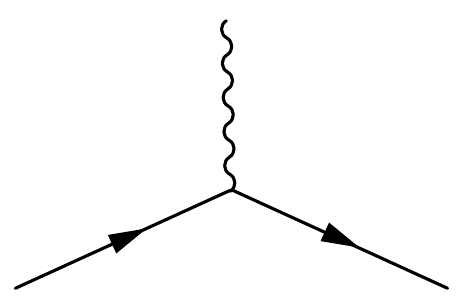}
        			\caption{The interaction vertex of electron with one ``photon'' field}
        			\label{eep}
\end{figure}

\begin{enumerate}

\item The electron propagator

\begin{gather}
\nonumber
\Braket{\eta(y) \bar{\eta}(x)} = M^{-1}_0(x, y) = \\ = \displaystyle \frac{1}{L_t L_x L_y} \sum\limits_{k^0, \vec{k}}
\frac{e^{i k (x - y)}}{(1 - e^{i k^0 \delta \tau} e^{ \delta \tau \mu} )^2 - \delta \tau^2 E^2(\vec{k}) e^{2 i k^0 \delta \tau} e^{2 \delta \tau \mu}} \times \nonumber \\ \times \begin{pmatrix}  1 - (1 + m c^2 \delta \tau) e^{i k^0 \delta \tau } e^{2 \delta \tau \mu} & \varphi(\vec{k}) \delta \tau e^{i k^0 \delta \tau } \\ \varphi^*(\vec{k}) \delta \tau e^{i k^0 \delta \tau } & 1 - (1 - m c^2 \delta \tau) e^{i k^0 \delta \tau } e^{ \delta \tau \mu} \end{pmatrix},
\label{el_prop}
\end{gather}

where $\varphi(\vec{k})$ was introduced in (\ref{phi}).

Note that the matrix $M^{-1}_0(x, y)$ is the $2 \times 2$ matrix in the ''sublattice space'',
\vspace{-0.25cm}

$$
M^{-1}_0(x, y) = \begin{pmatrix} M^{AA}_0(x, y) & M^{AB}_0(x, y) \\ M^{BA}_0(x, y) & M^{BB}_0(x, y) \end{pmatrix},
$$ 

its elements correspond to the propagation between one sublattice ($AA$ and $BB$) and different sublattices ($AB$ and $BA$).
\begin{figure}[H]
        \centering
                \includegraphics[width=0.25\textwidth]{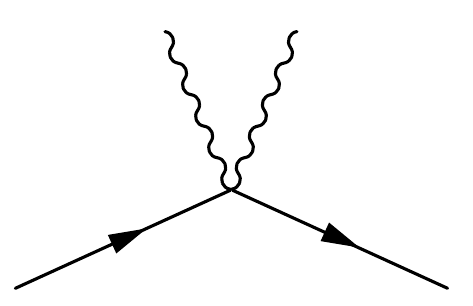}
        			\caption{The interaction vertex of electron with two ``photon'' fields}
        			\label{eepp}
\end{figure}
\item The hole propagator $\Braket{ \eta(y) \bar{\eta}(x)}$
can be obtained from the electron propagator with the substitution $\mu \to -\mu$.

\item The propagator of the ``photon'' field $\varphi$
\begin{gather*}
\Braket{\varphi(\vec{x}, x^0) \varphi(\vec{y}, y^0)} = \frac 1 {\delta \tau} \delta_{x^0, y^0} \hat{V}(\vec{x}, \vec{y}) =  \\ = \frac{1}{\delta \tau L_{\tau} L_x L_y} \sum\limits_{k^0, \vec{k}} e^{i k (x - y)}  \begin{pmatrix}\tilde{V}^{AA}(\vec{k}) & \tilde{V}^{AB}(\vec{k}) \\ \tilde{V}^{BA}(\vec{k}) & \tilde{V}^{BB}(\vec{k}) \end{pmatrix}.
\end{gather*}

\end{enumerate}

The potential is instantaneous and acts in one Euclidean time slice.

Using the expressions for the partition function, for the electron and hole propagators
one can show that in the limit $\delta \tau \to 0$ the charge of graphene sheet  $\langle Q \rangle$ is
\begin{gather}
\langle Q \rangle = -\sum_x \biggl ( \Tr (M_{xx}^{-1}) - \Tr ({\bar{M}}_{xx}^{-1}) \biggr ) =  \nonumber \\
= 2 \sum_{\vec k} \biggl ( \frac 1 {e^{\beta (E(\vec k) - \mu)} +1 } - \frac 1 {e^{\beta (E(\vec k) + \mu)} +1 } \biggr ),
\end{gather}
as it should be.

\subsection{Vertices}
The partition function (\ref{part_function}) takes into account the interactions between electrons/holes and ``photons''. The interaction vertices
of ``photons'' and electrons can be derived with an arbitrary accuracy through the expansion of the expression for the fermion operator:

\begin{widetext}

\begin{figure}[h!]

\begin{minipage}[t!]{0.45\textwidth}               
               	\includegraphics[width=0.6\textwidth]{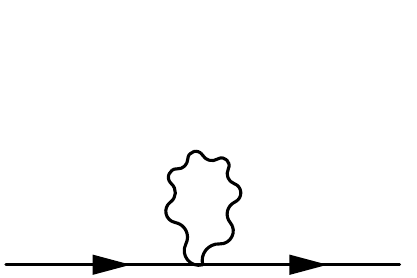}
                \caption{Diagram that contributes to the self energy $\Sigma_1(p)$}
                \label{diagT1}
\end{minipage}
\hfill
\begin{minipage}[t!]{0.45\textwidth}                
                \includegraphics[width=0.6\textwidth]{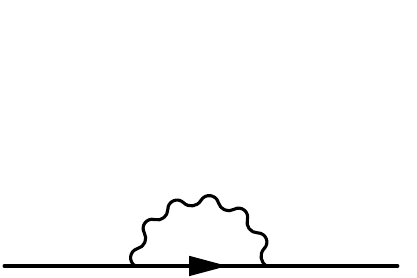}
                \caption{Fock diagram that contributes to the self energy $\Sigma_2(p)$}
                \label{diagT2}
\end{minipage}  
             
\end{figure} 

\begin{gather}
\nonumber
\mathcal{\delta M}_{l_1, l_2} (x, y) = \bigr ( e^{i \delta \tau \varphi_{l_1} (x)} -1 \bigl)~ e^{ \delta \tau \mu } \left(
  -  \delta_{\vec{x},\vec{y}} \delta_{l_1, l_2} \delta_{x^0 + \delta \tau, y^0}  + t \delta \tau \left[  \delta_{l_1, A}, \delta_{l_2, B} \sum\limits_{i = 1}^3 \delta_{x^0 + \delta \tau, y^0} \delta_{\vec{x} + \vec{\rho}_i, \vec{y}} + \right. \right. \\  \left. \left. +  \delta_{l_1, B} \delta_{l_2, A} \sum\limits_{i = 1}^3 \delta_{x^0 + \delta \tau, y^0} \delta_{\vec{x} - \vec{\rho}_i, \vec{y}} \right] + m c^2 \delta \tau \left[  \delta_{l_1, A} \delta_{l_2, A} \delta_{\vec{x}, \vec{y}} \delta_{x^0 + \delta \tau, y^0} -  \delta_{l_1, B} \delta_{l_2, B} \delta_{\vec{x}, \vec{y}} \delta_{x^0 + \delta \tau, y^0}\right] \right),
\label{deltaM}
\end{gather}
\end{widetext}

in powers of $\delta \tau$.  The simplest interaction vertex is shown in Fig. \ref{eep}. Using the formula (\ref{deltaM}), one can show that

at the leading order approximation in $\delta \tau$--expansion this vertex can be written as follows
\begin{gather}
V^{(3)}  = \bigr(-i \delta \tau \bigl) \sum_{l_1,l_2,x,y} \left(
\delta_{\vec{x},\vec{y}} \delta_{l_1, l_2} \delta_{x^0 + \delta \tau, y^0} \right) \bar \psi_{l_1}(x) \varphi_{l_1}(x) \psi_{l_2}(y)
\end{gather}

where $l_1, l_2$ are the sublattice indices.

The next vertex, which is required for the subsequent analysis and describes the interaction between two electrons and two ``photons'', is shown in Fig. \ref{eepp}.

At the leading order approximation it can be written as:
\begin{gather}
V^{(4)}  =   \frac {\delta \tau^2} 2  \sum_{l_1,l_2,x,y} \left(
\delta_{\vec{x},\vec{y}} \delta_{l_1, l_2} \delta_{x^0 + \delta \tau, y^0} \right) \bar \psi_{l_1}(x) \varphi^2_{l_1}(x) \psi_{l_2}(y)
\end{gather}

There are also additional vertices, coupling electrons to $3, 4 \dots, n, \dots$ ``photons'' (because the $\varphi$ field stands in the exponent).
However, they are suppressed by the additional factors $\delta \tau$ and give no contribution to the final answer in the limit $\delta \tau \to 0$. Note also that
in this section we presented only vertices with electrons. The vertices with holes can be found similarly.

\section{One-loop  corrections to the electron propagator}
First we are going to consider one-loop corrections at zero chemical potential and nonzero temperature.
One-loop corrections to the electron propagator can be expressed in terms of the self-energy function $\Sigma(p)$:
\begin{gather}
M^{-1}(p) = \frac{1}{M_0(p) - \Sigma(p)}.
\end{gather}
At the leading order approximation there are two diagrams shown in Fig. \ref{diagT1}, \ref{diagT2} that contribute to $\Sigma(p)$:
\begin{gather*}
\Sigma_1(p) = +\displaystyle \frac{1}{2} \delta \tau \begin{pmatrix} V_{00} & 0 \\ 0 & V_{00} \end{pmatrix} e^{i p^0 \delta \tau},\\
\Sigma_2(p) = -\displaystyle \Sigma_1(p) + \frac{e^{i p^0 \delta \tau} \delta \tau}{2 L_x L_y} \sum \limits_{\vec{k}} \tanh \left(\frac{E(\vec{k})}{2 T} \right)  \times \nonumber \\ \times  \begin{pmatrix} - \frac{m c^2}{E(\vec{k})}\tilde{V}^{AA}(\vec{p} - \vec{k}) &  \frac{\varphi(\vec{k})}{E(\vec{k})}\tilde{V}^{AB}(\vec{p} - \vec{k}) \\ \frac{\varphi^*(\vec{k})}{E(\vec{k})}\tilde{V}^{BA}(\vec{p} - \vec{k}) & \frac{m c^2}{E(\vec{k})}\tilde{V}^{BB}(\vec{p} - \vec{k}) \end{pmatrix}.
\end{gather*}

\begin{figure*}[t!]
        \centering
        \begin{subfigure}[b]{0.6\textwidth}
                \includegraphics[width=\textwidth]{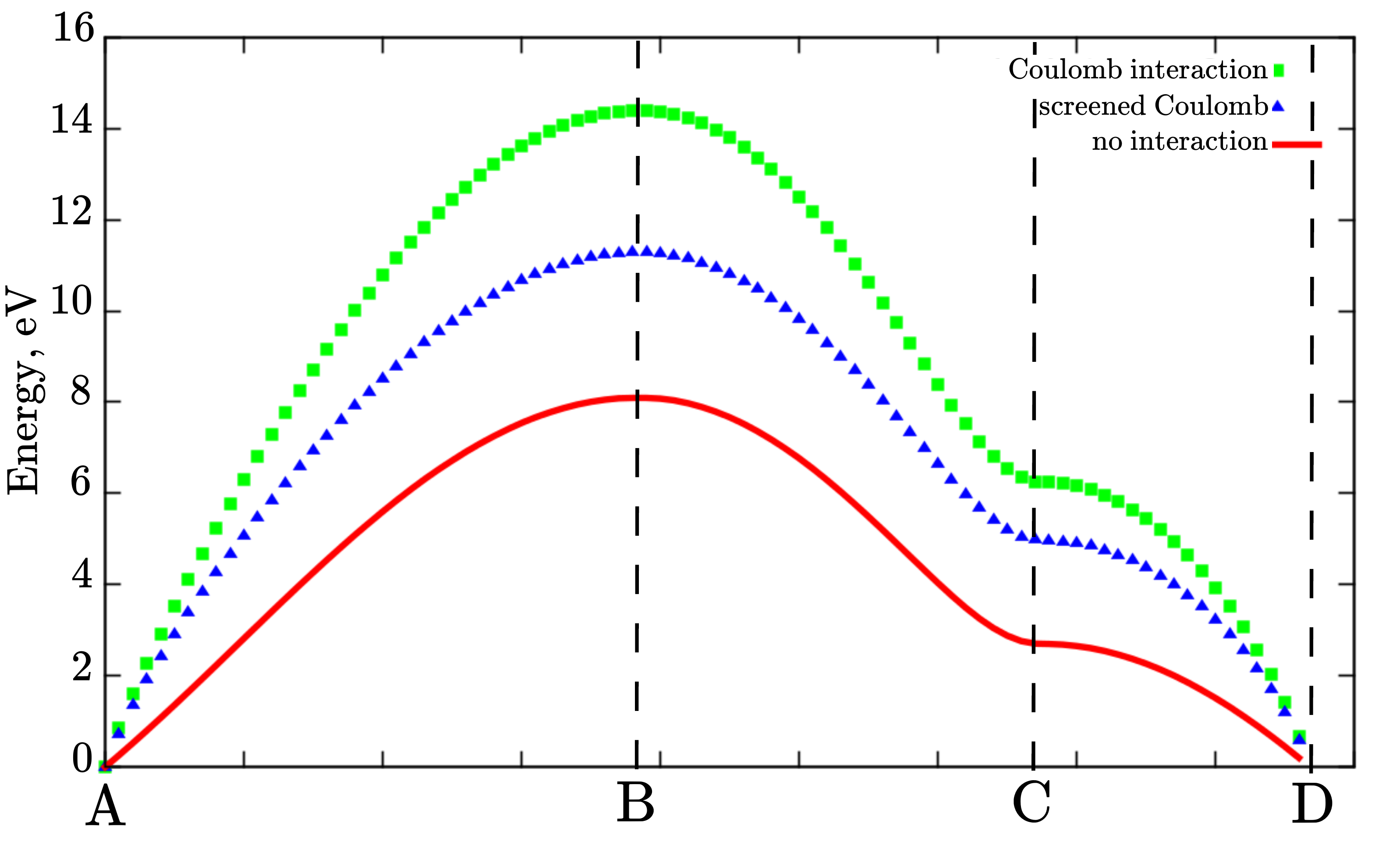}
                \subcaption{Free and renormalized energy $E(\vec{k})$ profile.}
        \end{subfigure}
        \qquad
        \begin{subfigure}[b]{0.3\textwidth}
                \includegraphics[width=\textwidth]{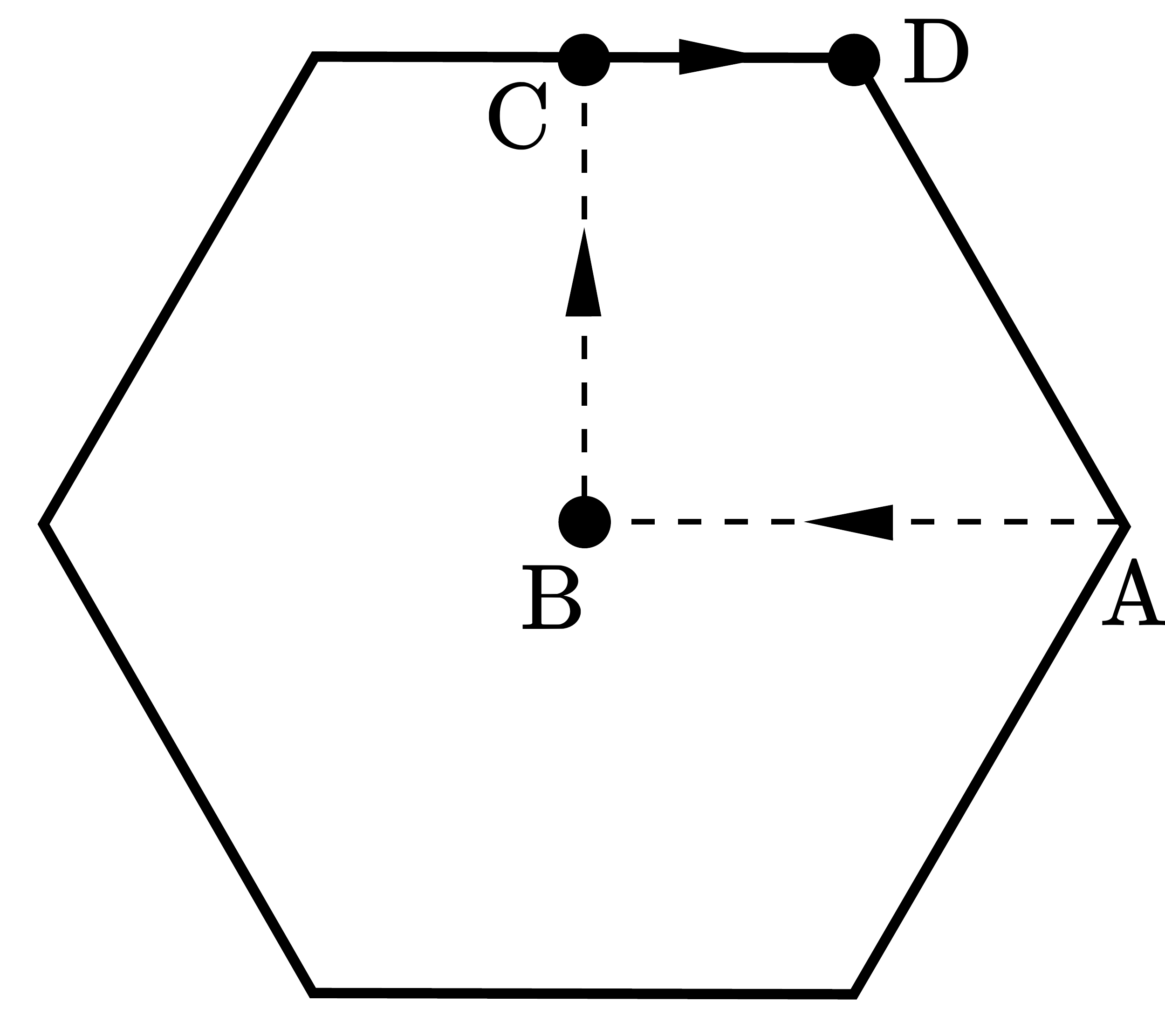}
                \subcaption{The spectrum profile is drawn on the polygon $ABCD$.}
        \end{subfigure}
		\caption{Energy spectrum profile of electrons with Coulomb, screened Coulomb \cite{Wehling} interactions and without interactions. The calculation was carried out at $T = 0.1$ eV at infinitely large lattice.}
                        \label{energy}
\end{figure*}

The inverse of the free electron propagator $M_0(p)$ can be written in the following form (see formula (\ref{el_prop})):
\begin{gather}
\label{m0}
M_0(p)  =
\begin{pmatrix}  1 -  e^{i p^0 \delta \tau } & 0 \\ 0 & 1 -  e^{i p^0 \delta \tau} \end{pmatrix} +
e^{i p^0 \delta \tau } \delta \tau \begin{pmatrix}   m c^2  & \varphi(\vec{p})
\\ \varphi^*(\vec{p})  & - m c^2  \end{pmatrix},
\end{gather}
The final expression for $\Sigma(p)$ is

\begin{gather}
\Sigma(p) = \Sigma_1(p) + \Sigma_2(p) = \frac{e^{i p^0 \delta \tau} \delta \tau}{2 L_x L_y} \sum \limits_{\vec{k}} \tanh \left(\frac{E(\vec{k})}{2 T} \right) \times \nonumber\\ \times \begin{pmatrix} - \frac{m c^2}{E(\vec{k})}\tilde{V}^{AA}(\vec{p} - \vec{k}) &  \frac{\varphi(\vec{k})}{E(\vec{k})}\tilde{V}^{AB}(\vec{p} - \vec{k}) \\ \frac{\varphi^*(\vec{k})}{E(\vec{k})}\tilde{V}^{BA}(\vec{p} - \vec{k}) & \frac{m c^2}{E(\vec{k})}\tilde{V}^{BB}(\vec{p} - \vec{k}) \end{pmatrix}.
\label{sigma_1l}
\end{gather}

From the formula (\ref{sigma_1l}) one can see that one-loop corrections are reduced to the renormalization
of the mass $m$ and the function $\varphi(\vec p)$. The renormalized mass now depends on the point in the Brillouin zone
and has the form
\begin{gather}
m^R(\vec{p}) = m + \frac{m}{2 L_x L_y} \sum \limits_{\vec{k}} \tanh \left(\frac{E(\vec{k})}{2 T} \right) \times \nonumber\\ \times \frac{1}{E(\vec{k})}\tilde{V}^{AA}(\vec{p} - \vec{k}).
\label{mass_r}
\end{gather}
The expression for the renormalized function $\varphi^R(\vec{p})$ is
\begin{gather}
\varphi^R(\vec{p}) = \varphi(\vec{p}) + \frac{1}{2 L_x L_y} \sum \limits_{\vec{k}} \tanh \left(\frac{E(\vec{k})}{2 T} \right) \times \nonumber \\ \times \frac{\varphi(\vec{k})}{E(\vec{k})}\tilde{V}^{AB}(\vec{p} - \vec{k}).
\label{phi_r}
\end{gather}
Thus one-loop corrections conserve the form of the propagator (\ref{el_prop}) and
substitute the free mass $m$ and the function $\varphi(\vec{p})$ with the renormalized expressions (\ref{mass_r}), (\ref{phi_r}).
From this one can conclude that the energy spectrum of the quasiparticles at the one-loop approximation is  $E^2 = (m^R c^2)^2 + (\varphi^R)^2$.
In order to estimate the size of one-loop corrections in Fig. \ref{energy} we plot the energy spectrum profile of
electron with Coulomb, screened Coulomb \cite{Wehling} interactions and without interactions. The calculation
was carried out at $T = 0.1\, \mbox{eV}$ at infinitely large lattice.

The formulas (\ref{mass_r}) and (\ref{phi_r}) can be used to reproduce well-known results of the effective theory
of graphene and generalize them to the case of nonzero temperature. To this end, we consider large lattice $L_x,\; L_y \to \infty$ with the Coulomb interactions between electrons near the Fermi point. Then the formulas (\ref{mass_r}) and (\ref{phi_r}) can be written as
\begin{gather}
m^R(\vec{p} = \vec{p}_{F}) = \nonumber \\ =  m \left( 1 + \frac{1}{2} \frac{\alpha}{ (v_F/c)} \left[\log \left(\frac{v_F \Lambda}{2 c T} \right) + \gamma - \log \pi/4 + \mathcal{O}(\Lambda^{-1})\right] \right), \nonumber
\\
v^R_F = \nonumber \\ = v_F \left( 1 + \frac{1}{4} \frac {\alpha} {(v_F/c)} \left[\log \left(\frac{v_F \Lambda}{2 c T} \right) + \gamma - \log \pi/4 + \mathcal{O}(\Lambda^{-1})\right]
\right ).
\label{renorm}
\end{gather}
where $\Lambda$ is the ultraviolet cut-off, $\gamma \approx 0.577...$ is the Euler's constant. These formulas are in agreement with
the predictions of effective theory of graphene at the leading logarithmic accuracy \cite{Gonzalez:1993uz, Kotov:1}.

\begin{figure*}[t!]
\begin{minipage}[t!]{0.3\textwidth} 
\centering
                \includegraphics[width=0.68\textwidth]{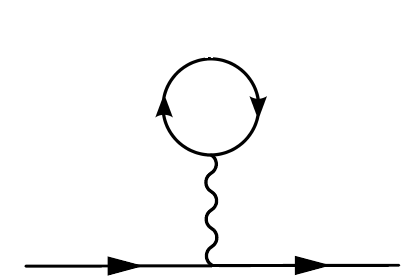}
				\caption{Hartree diagram that contributes to the one-loop electron propagator at $\mu \neq 0$.}
        			\label{diagmu}
\end{minipage}
\qquad
\qquad
\qquad
\qquad
\qquad 
\qquad
\begin{minipage}[t!]{0.3\textwidth} 
\centering
		\includegraphics[width=0.6\textwidth]{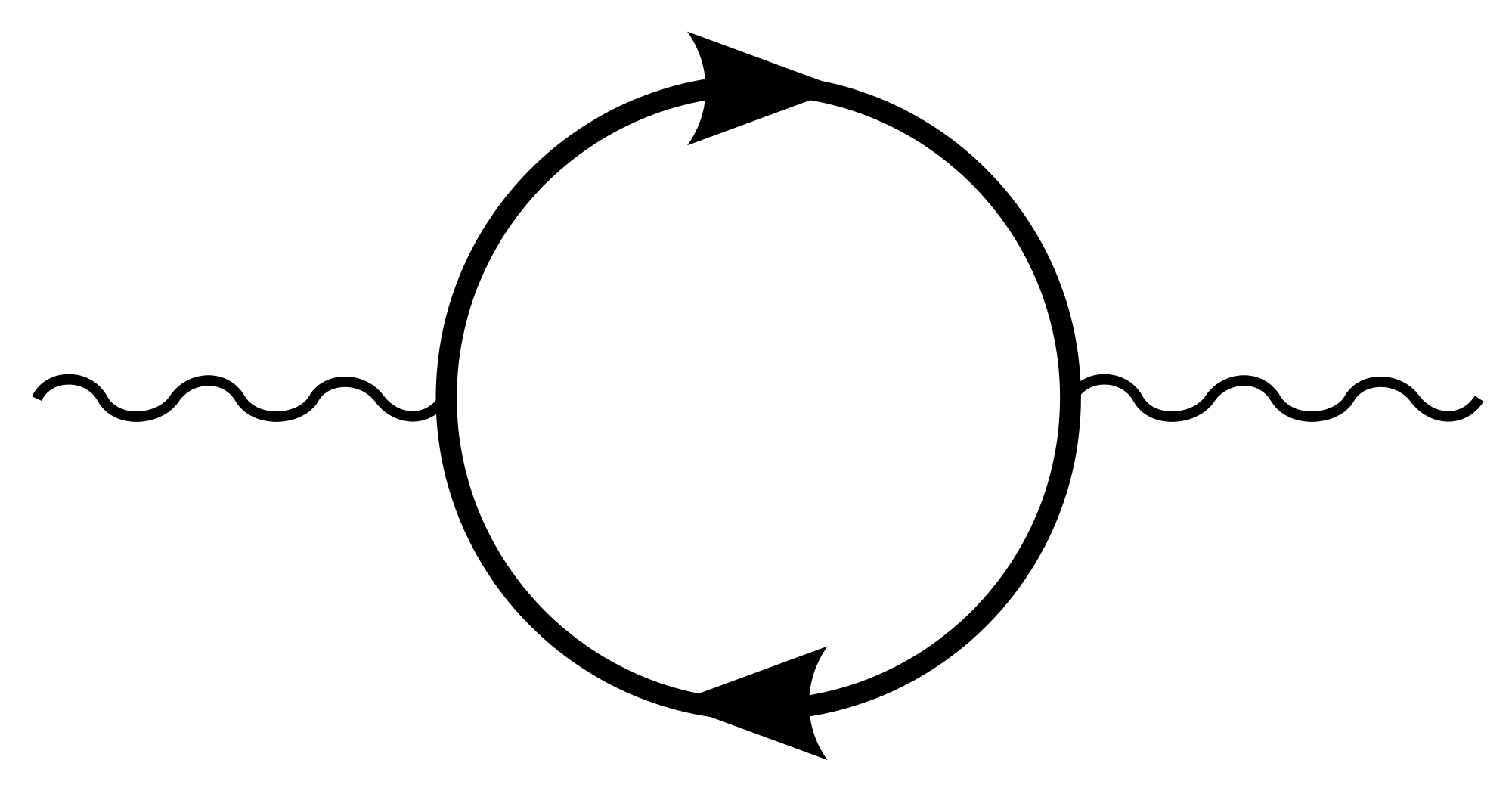}
        \caption{Empty-loop diagram that contributes to the renormalization of the interaction potential.}
        \label{diagP}
\end{minipage}        
\end{figure*}	

During the derivation of the formulas (\ref{renorm})
we carried out the integration over the two-dimensional sphere of radius $\Lambda$. It was also assumed that $m c^2 \ll T$,
$\Lambda$ is much larger than all other energy scales involved. Note that the theory is regularized by the temperature in the infrared region.
Note also that the ratio of the ultraviolet and the infrared cut-offs $\Lambda/T$ under the logarithm is multiplied by the $v_f / c$
that considerably reduces the total renormalization of the Fermi velocity.

In order to study the finite density effects we consider one-loop corrections at zero temperature and nonzero chemical potential.
In addition to the diagrams shown in Figures \ref{diagT1}, \ref{diagT2} there is a contribution of the interaction of the
electron propagator with vacuum represented by the diagram shown in Fig. \ref{diagmu}. As before, one-loop corrections at nonzero $\mu$
lead to the renormalization of the parameters of free propagator preserving its structure. In particular,
the mass and the function $\varphi$ at one-loop approximation can be written as follows

\begin{gather}
m^R(\vec{p}) = m + \frac{m}{2 L_x L_y} \sum \limits_{\vec{k}}  \frac{1}{E(\vec{k})}\tilde{V}^{AA}(\vec{p} - \vec{k})
~\theta (E(\vec k)-\mu), \nonumber \\
\varphi^R(\vec{p}) = \varphi(\vec{p}) + \frac{1}{2 L_x L_y} \sum \limits_{\vec{k}} \frac{\varphi(\vec{k})}{E(\vec{k})}\tilde{V}^{AB}(\vec{p} - \vec{k})
~\theta (E(\vec k)-\mu).
\label{renmu}
\end{gather}
In the limit of large lattice $L_x,\; L_y \to \infty$, Coulomb interaction and the linear electron spectrum near the Fermi point, the formulas (\ref{renmu}) can be written as
\begin{gather}
m^R(\vec{p} = \vec{p}_{F})  =  m \left( 1 + \frac{1}{2} \frac{\alpha}{(v_F / c)} \log \left[\frac{v_F \Lambda}{c \mu} \right] \right), \nonumber
\\
v^R_F = v_F \left( 1 + \frac{1}{4} \frac {\alpha} {(v_F / c)} \log \left[ \frac{v_F \Lambda}{c \mu}  \right]
\right ).
\label{renormmu}
\end{gather}
Similarly to the case of nonzero temperature chemical potential $\mu$ plays the role of the infrared cut-off.
It should be noted now that there are four scales that can play the role of the infrared cut-off: the fermion mass $m$, the
chemical potential, the temperature and the inverse lattice size. We believe that in the infrared limit the theory
is regularized by the largest of these scales. Note that the formulas (\ref{renorm}), (\ref{renmu}) take these
effects into account exactly.

Note also that the formulas (\ref{renorm}) can be written in the form (\ref{renormmu}) if instead of $\Lambda$
one uses new ultraviolet cut off $\Lambda'=2 e^{\gamma}/\pi \cdot \Lambda \sim 1.13 \cdot \Lambda$ and substitute the
chemical potential by the temperature.

For this reason at temperature equal to chemical potential the renormalization due to temperature 
effects is a little bit larger than the renormalization due to nonzero chemical potential. 
In this consideration we did not take the Debye screening of the interaction potential at large distances
and the screening of the Coulomb potential at small distances into account, that will be done in the last section.
$ $\\
$ $\\

\section{One-loop corrections to ``photon'' propagator and dielectric permittivity of graphene}

In this section we calculate one-loop correction to the instantaneous ``photon'' propagator: $\Braket{\varphi(\vec{x}, x^0) \varphi(\vec{y}, y^0)}$.

One-loop corrections to the propagator in the momentum space can be expressed in terms of the polarisation operator
$\mathcal{\hat{P}} = \begin{pmatrix}\mathcal{P}^{AA} & \mathcal{P}^{AB} \\ \mathcal{P}^{BA} & \mathcal{P}^{BB} \end{pmatrix},$
as
\begin{gather}
{\tilde V}^{R}(\vec k) = \frac 1 {\hat \epsilon(\vec k)} \times {\tilde V}(\vec k)  \nonumber \\
\hat \epsilon(\vec k) = 1 - {\tilde V}(\vec k) \times \mathcal{\hat P}(\vec k),
\label{epsilonR}
\end{gather}

where ${\tilde V}^{R}(\vec k), {\tilde V}(\vec k)$ are renormalized and tree level Fourier transforms of the potential, which
are the $2 \times 2$ matrices in the sublattice space.  In the limit of $\delta \tau \to 0$ and one-loop approximation the only  diagram shown in Fig. \ref{diagP} contributes.

The expression for the polarization operator at one-loop approximation can be written in the following form:

\begin{widetext}
\begin{gather}
\mathcal{P}^{AA}(\vec{k}) = \mathcal{P}^{BB}(\vec{k}) =
\frac{1}{L_x L_y} \sum\limits_{\vec{p}} \frac 1 {E^2(\vec p- \vec k) - E^2(\vec p)} \biggr (
\frac {m^2 c^4 + E^2(\vec p)} {E(\vec p)} \tanh \frac{E(\vec p)}{2 T} - \frac {m^2 c^4 + E^2(\vec p - \vec k)} {E(\vec p - \vec k)} \tanh \frac{E(\vec p -\vec k)}{2 T} \biggl ),  \nonumber
 \\
\mathcal{P}^{AB}(\vec{k}) = \mathcal{P^*}^{BA}(\vec{k}) = \frac{1}{L_x L_y} \sum\limits_{\vec{p}} \frac {\varphi(\vec p) \varphi^*(\vec p- \vec k)} {E^2(\vec p- \vec k) - E^2(\vec p)} \biggr (
\frac {1} {E(\vec p)} \tanh \frac{E(\vec p)}{2 T} - \frac {1} {E(\vec p - \vec k)} \tanh \frac{E(\vec p -\vec k)}{2 T} \biggl ).
\label{polarization}
\end{gather}
\end{widetext}
Using formulas (\ref{epsilonR}), (\ref{polarization}) one can show that at large distances and at small temperature
the expression for the interaction potentials for all the sublattice indices take the form
\begin{gather}
{\tilde V}^{R}(\vec k) = \frac{2 \pi \alpha c}{\epsilon_0}   \frac {1}{(|\vec{k}| + m_D c)},
\label{potential}
\end{gather}
where
\begin{gather}
\label{epsilon}
\epsilon_0 = 1 + \frac{\pi}{2} \frac {\alpha} {(v_F / c)},  \\ \nonumber
m_D = \frac{8 \alpha \log 2}{\epsilon_0 v_F^2} T.
\end{gather}
Now one may see that at zero temperature the interaction potential is the Coulomb potential screened by the dielectric permittivity $\epsilon_0$,
which is in agreement with the RPA result \cite{Katsnelson}. At nonzero temperature there is a two-dimensional Debye screening of the Coulomb potential with the Debye mass $m_D$, which agrees with the results of the paper \cite{Braguta:2013klm}.

Similarly one can study the question how the nonzero density acts on the dielectric permittivity of graphene. To get the expression for
the polarization operator in this case one can use the formulas (\ref{polarization}) and carry out the following substitution
$\tanh (E/2T) \to \theta (E - \mu)$. It is not difficult to find out that at large distance the expression for the interaction potential 
has the form (\ref{potential}) with $\epsilon$ given by the equation (\ref{epsilon}) and the Debye mass
\begin{gather}
m_D = \frac {4 \alpha} {\epsilon_0 v_F^2}  |\mu|.
\label{Debyemu}
\end{gather}
The last expression agrees with the RPA result \cite{Katsnelson}.

At the end of this section we plot the dielectric permittivity of graphene $\epsilon$ obtained from formulas (\ref{epsilonR}), (\ref{polarization})
as a function of distance $r$ in units of the lattice spacing $a$ (Fig. \ref{eps}).
The calculation was carried out for suspended graphene with the interaction potential from paper \cite{Ulybyshev:2013swa}
and for different external conditions: $T = 0, \mu=0$, $T = 0, \mu = 0.026$ eV ($n \sim 6.5\cdot10^{10}$ cm$^{-2}$) and $T=0.026$ eV,  $\mu=0$.
In order to compare our results with RPA we plot the dielectric permittivity given by formula (\ref{epsilon}).
For small distance we compared our results with the results of Monte-Carlo simulation \cite{Astrakhantsev}. It is seen
that the Monte-Carlo results are in a good agreement with one-loop results.

From Fig. \ref{eps} it is seen
 that at zero temperature the $\epsilon$ starting from the value $\sim 2$ at $r = 0$ approaches the value $\epsilon_0$ (\ref{epsilon})
already for $r/a > 5$. At nonzero temperature and chemical potential the larger the $r$ the larger the $\epsilon$. This behaviour
can be attributed to the Debye screening effect.

\section{Discussion and conclusion}

This paper is devoted to the perturbation theory which can be used for studying the properties of graphene at finite temperature and nonzero
chemical potential.
This perturbation theory is based on the tight-binding Hamiltonian on hexagonal lattice and arbitrary interaction potential between electrons, which
is considered as a perturbation. We built the partition function for this theory, derived Feynman rules and expressions for free propagators.

As an example of the application we calculated one-loop corrections to the electron propagator. It was shown that
one-loop corrections lead to the renormalization of the bare mass and the function $\varphi(\vec k)$ conserving the
structure of the propagator. Using this result we calculated one-loop energy spectrum of electrons,
renormalized fermion mass and Fermi velocity. 

\begin{figure*}[t!]
\begin{minipage}[t!]{0.49\textwidth}
                \includegraphics[width=1.0\textwidth]{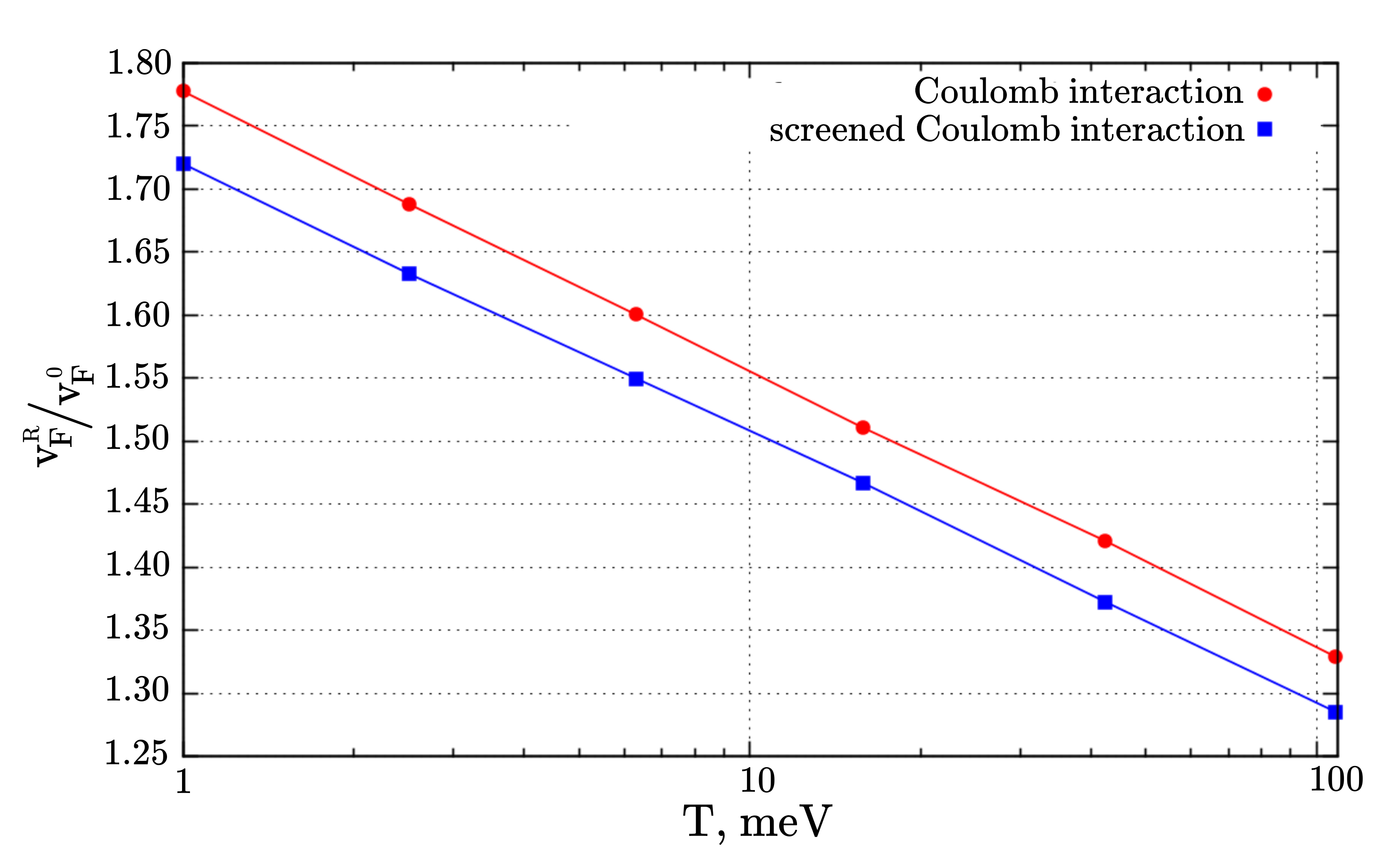}
                \caption{The renormalization factor for the Fermi velocity as a function of the temperature for graphene
on hBN for Coulomb and screened at small distances Coulomb interactions \cite{Wehling}.}
 \label{renvf}
 \end{minipage}
 \hfill
 \begin{minipage}[t!]{0.49\textwidth}
 \centering
                \includegraphics[width=1.0\textwidth]{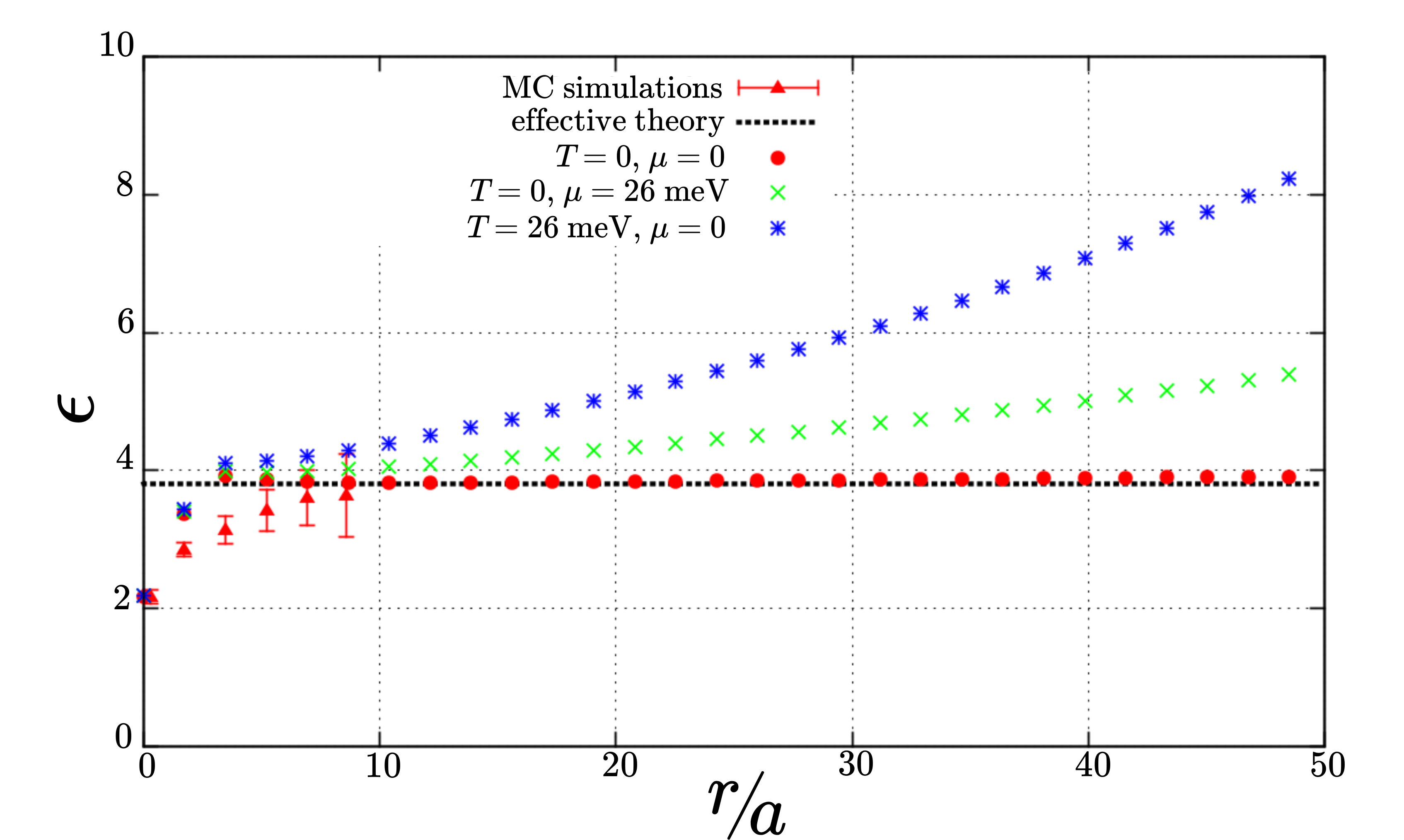}
\caption{The dielectric permittivity of suspended graphene $\epsilon$ as a function of distance $r$ in units of distance between the neighbouring carbon atoms $a$ for different external conditions.
The dotted line corresponds to the RPA expression (\ref{epsilon}). The points shown as
red triangles are the results of Monte-Carlo simulation.
}
 \label{eps}
 \end{minipage}
\end{figure*}

In order to estimate the value of the renormalization,
in Fig. \ref{renvf} we plotted the renormalization factor for the Fermi velocity as a function of the temperature for graphene
on hBN for the Coulomb and screened at small distances Coulomb interactions \cite{Wehling}. During the calculation we took the screening
of the potentials by the dielectric permittivity (\ref{epsilonR}) into account. The results can be well described by the formula
\begin{gather}
v_F^R(T)/v_F^0 =  \biggl ( 1 + A \cdot \log \biggr [ \frac  {\Lambda} {\mu} \biggl ]  \biggr )
\label{vfT}
\end{gather}
where for the Coulomb interaction $A=0.096, \Lambda=3.2~\mbox{eV}$ and for the screened Coulomb interaction $A=0.093, \Lambda=2.4~\mbox{eV}$. 
The effective ultraviolet cut-off $\Lambda$ is sensitive to the values of the potential at small distances, contrary to the coefficient $A$.
Note also that that the coefficient $A$ is well described by the expression
$A \approx \alpha c / (4 v_F \epsilon') = 0.094,~~\epsilon'=(\epsilon_{BN}+1)/2 + \epsilon_0-1=6.7$.

In addition we considered one-loop corrections to the electron propagator at zero temperature and nonzero chemical potential.
It was shown that similarly to the nonzero temperature case one-loop corrections conserve the structure of the
propagator leading to the renormalization of already existing parameters. In particular, we obtained the formulas
for the renormalization of the Fermi velocity and the fermion mass in this case.

In paper \cite{PNAS} the renormalization of the Fermi velocity was studied through the measurement of quantum
capacitance at nonzero chemical potential. In this study the graphene layer was placed inside hBN, which reduced
the strength of the interaction between electrons. This allows us to expect that perturbation theory works well
in this case and we can compare our results with the the result of \cite{PNAS}. The results of the measurements
of the Fermi velocity can be well fitted by the formula
\begin{gather}
v_F(\mu) = v_F (\mu_0) \biggl ( 1 + \frac 1 4 \frac {\alpha} {\epsilon (v_F / c)} \log \biggr [ \frac  {\mu_0} {\mu} \biggl ]  \biggr ) = \nonumber \\
=v_F (\mu_0) \biggl ( 1 + A~ \log \biggr [ \frac  {\mu_0} {\mu} \biggl ]  \biggr ),
\label{vf}
\end{gather}
where $v_F(\mu_0)=0.85 \times 10^{6}\, m/s,~\mu_0 = 3.2\, \mbox{eV}~ (n_0 = 10^{15}~ \mbox{cm}^{-2})$,
 $\epsilon \simeq 8$ and $A = 0.081$. Note
that the original formula for the Fermi velocity from \cite{PNAS} depends on the density $n$. In (\ref{vf}) we turn the
dependence on $n$ to the dependence on chemical potential.

  In the calculation we used the formulas (\ref{renmu}) with the interaction potential screened by the dielectric permittivity
(\ref{epsilonR}). The interaction potential at small distances was taken from the paper \cite{Wehling} and divided by
$\epsilon_{BN} \simeq 4.5$. At large distances we took the Coulomb potential screened by $\epsilon_{BN}$.
The results of the calculations can be well fitted by formula (\ref{vf}) with the parameters:
$\mu_0 = 2.9\, \mbox{eV}, A = 0.072$. This values are in reasonable agreement with  paper \cite{PNAS}.
In addition we carried out the calculation of the parameters $\mu_0, A$ for the Coulomb potential
screened by $\epsilon_{BN}$. Our result is $\mu_0 = 5.2~\mbox{eV}, A = 0.077$. Again 
we see that the value of the constant $\mu_0$ is sensitive to the values of the potential at 
small distances.

It is also interesting to compare the renormalization of the Fermi velocity 
due to nonzero $\mu$ and nonzero temperature. 
To this end we calculated nonzero temperature renormalization of the graphene layer 
placed inside the hBN for the Coulomb and screened at small distances Coulomb interactions. 
The results can be described by formula (\ref{vfT}) with the parameters: 
for the Coulomb interaction $A=0.075, \Lambda=4.5~\mbox{eV}$ and for the screened Coulomb interaction 
$A=0.073, \Lambda=3.05~\mbox{eV}$.

The other example of application of the perturbation theory is the calculation of one-loop corrections
to the interaction potential done in the previous section. We derived the one-loop expression for the dielectric permittivity
at nonzero temperature, nonzero chemical potential and the arbitrary interaction potential.

It is well known that the electrons in graphene form a strongly interacting system. 
So it is reasonable to consider the question how our results are affected 
by the higher order corrections. The authors of paper \cite{dassarma} considered the next-to-leading order (NLO) corrections 
to the Fermi velocity within effective theory of graphene. The main result of 
this paper in the statement that if one expands the Fermi velocity renormalization in 
the one-loop RPA potential instead of the usual Coulomb potential, 
the NLO corrections to the leading order(LO) result (\ref{renormmu}) turn out to be small. 
This allows us to expect a good accuracy of the formulas (\ref{mass_r}), (\ref{phi_r}), (\ref{renmu}) with one-loop potential given by the formulas (\ref{epsilonR}),
(\ref{polarization}) even for suspended graphene. 

The authors of the paper \cite{fogler} considered the NLO corrections to the polarization operator within the effective 
theory of graphene. The NLO value of the dielectric permittivity for suspended graphene is approximately by 30 \%
smaller than the LO result, which is not very large. Moreover, the Monte-Carlo results \cite{Astrakhantsev}
tell us that the higher order corrections to the LO result can be even smaller than 30 \%. For 
this reason one can expect that the accuracy of formulas (\ref{epsilonR}), (\ref{polarization}) is rather good.

\section*{Acknowledgements}

We would like to thank M. Ulybyshev and  A. Nikolaev for interesting discussions.
MIK acknowledges financial support from NWO via Spinoza Prize.
The work of VVB and NYB was supported by the Far Eastern Federal University,
by RFBR grants 14-02-01261-a, 15-02-07596, 15-32-21117 and the Dynasty Foundation.

\end{document}